\documentclass{article}
\usepackage{axodraw}
\usepackage{epsfig}
\usepackage{a4wide}
\begin{document}
\newcommand{\be}{\begin{equation}}
\newcommand{\ee}{\end{equation}}
\newcommand{\ba}{\begin{eqnarray}}
\newcommand{\ea}{\end{eqnarray}}
\begin{flushright}\rm LU TP 99-41\\hep-ph/9912398\\Revised February 2000
		   \end{flushright}
\begin{center}
{\Large\bf
Low Energy Constants from $K_{\ell 4}$ Form-Factors}\\[1cm]

{\bf G. Amor\'os$^{a,b}$, J. Bijnens$^b$ and P. Talavera$^b$}\\[0.5cm]
{$^a$ Department of Physics, P.O. Box 9
FIN-00014 University of Helsinki, Finland\\
$^b$ Department of Theoretical Physics 2, Lund University,
S\"olvegatan 14A, S22362 Lund, Sweden} 

\begin{abstract}
We have calculated the form-factors $F$ and $G$ in
$K\to\pi\pi\ell\nu$ decays ($K_{\ell 4}$) to two-loop order in Chiral
Perturbation Theory (ChPT). Combining this together with earlier two-loop
calculations an updated set of values for the $L_i^r$, the ChPT
constants at ${\cal O}(p^4)$, is obtained. We discuss the
uncertainties in the determination and the changes compared to previous
estimates.
\end{abstract}

{{\bf PACS numbers:}  13.20.Eb, 11.30.Rd, 12.39.Fe, 12.38.Aw}
\end{center}

{\bf 1.}~The theoretical study of $K \to \pi \pi l \nu$ decays,
$K_{\ell 4}$ decays, provides an interesting possibility to test our
understanding of the long-distance dynamics of the strong sector in
the Standard Model.
Perturbative QCD can not be applied
at energies lower than the spontaneous symmetry scale, 
$\Lambda_\chi \sim m_\rho$. One
has to resort to an effective approach to the full theory, 
Chiral Perturbation Theory (ChPT) \cite{GL1}, to
obtain a reliable model independent description of physical processes.
The main purpose in the study of $K_{\ell 4}$ decays is twofold: 
$i)$ This decay is one of the cleanest sources of $\pi\pi$ pairs at
low-energies and thus provides us with the possibility to check
$\pi\pi$-scattering near threshold.
$ii)$ The form-factors themselves are also directly of interest.
They provide a direct test of our understanding of the three-flavour sector
and are in addition one of the major inputs to determine the needed
constants to predict other quantities.
The early history can be found in the review \cite{Chounet:1972yy}.
$K_{\ell 4}$ have been treated in the context of ChPT at one-loop
\cite{Kl4oneloop} and in a dispersively improved one-loop representation
\cite{BCG}. These studies delivered the standard values
of the low-energy constants $L_1^r, L_2^r$ and $L_3^r$.

On the other hand,
$\pi\pi$-scattering has been studied at two-loop order in two-flavour
ChPT, direct \cite{pipi} and via dispersive methods \cite{Knecht}.
The main remaining uncertainty on the prediction of
$\pi\pi$-scattering is the value of the low-energy
constants. In \cite{pipi} the standard values from \cite{BCG} were used.
A Roy equation
analysis  using high energy $\pi\pi$-data, gave significantly different values
for these constants \cite{Girlanda:1997ed}. There are several
possible sources for this discrepancy:
the Omn\`es representation of \cite{BCG} is not sufficient for
$K_{\ell 4}$; the presence of large systematic effects in
the $\pi\pi$-scattering
data base and Roy analysis; or large corrections to the one-loop
relation between two- and three-flavour ChPT constants.
Therefore, and since new experiments are underway, a full two-loop calculation
of $K_{\ell 4}$ is necessary.
A first step was the estimate of the double logarithm corrections of
${\cal O}(p^6)$  to $K_{\ell 3}$
and $K_{\ell 4}$ \cite{BCEdble}. It was shown there they could be large.
In addition pions in the $I=0$, S-wave have strong final state
interactions.
We describe here first results of the full $K_{\ell 4}$ ${\cal O}(p^6)$
calculation.\\

{\bf 2.} $K_{\ell 4}$ decays are described by the matrix element
\be
\label{matrix}
M^{ij}  = \frac{G_F}{\sqrt{2}} V_{us}^*\,\bar{u}(p_\nu)
\gamma_\mu(1-\gamma_5) v(p_{\ell}) 
\langle\pi^i(p_i)\pi^j(p_j)\vert V^\mu-A^\mu\vert K(k)\rangle \,.
\ee
Lorentz invariance
allows to parametrize the hadronic part by four form-factors 
\ba
\label{formfactors}
&&\langle\pi^i\pi^j\vert A^\mu\vert K\rangle\ = - \frac{i}{m_K} \left[
P^\mu F^{ij} + Q^\mu G^{ij} 
+ L^\mu R^{ij} \right]\,, \nonumber\\&&
\langle\pi^i\pi^j\vert V^\mu\vert K\rangle\ = - \frac{1}{m_K^3} 
\epsilon^{\mu\nu\alpha\beta} L_\nu P_\alpha Q_\beta H^{ij}\,,
\ea
with $P=p_i + p_j, Q = p_i - p_j$ and $L = p_{\ell}+p_\nu$.
$p_i$ is the momentum of  $\pi^i$. 
All four form-factors are
dimensionless functions of three variables, 
$s_\pi=P\cdot P, u_\pi=(k-p_i)^2$ and $t_\pi = (k-p_j)^2$.
They can be decomposed,
into a part
symmetric or antisymmetric under $t_\pi\leftrightarrow u_\pi$ interchange,
corresponding respectively to the isospin 0 or 1 part for $F,R$ and
isospin 1 or 0
for $G,H$.
These processes are dominated by $F$ and $G$.
$H$ has a fairly small influence and a
one-loop calculation, as performed in \cite{Hanomaly}, is sufficient to
reach ${\cal O}(p^6)$ since it comes through the
odd-intrinsic-parity sector.
$R$ only contributes proportional to $m_\ell^2$ and
can thus be neglected in the decays with an electron. 

In the remainder we concentrate on the $ij=+-$ channel.
The others can be derived from it using isospin relations.
The older available experiments are compatible with the
 $K^+\to\pi^+\pi^- e^+\nu_e$ experiment \cite{Rosselet}.
The most recent other measurement is $K_L\to\pi^\pm\pi^0e^\mp\nu$
\cite{Makoff}.

The experiment \cite{Rosselet}
relied on a partial wave analysis of the form-factors.
The analysis only kept $s$ and $p$ waves, the effect of
$d$-waves as well as of $s_\ell$ was within the measurement
errors and neglected in the further analysis.
They used the parametrization
\be
\label{partial}
F = f_s e^{i\delta_s} + f_p e^{i\delta_p}\cos\theta_\pi
+ \mbox{{D-wave}}+\ldots\,,
\quad
G =  g e^{i\delta_p}+ \mbox{{D-wave}} + \ldots\,,
\ee
where $\delta_i$ are the i-wave strong two-pion final state phase shifts, and
$f_s, f_p$ and $g$ are defined to be real functions with
argument $s_\pi$. They then divided the data in $5$ energy bins,
observed an $s_\pi$ dependence in $f_s$ and one compatible with it
for $g$. The final fit was performed using
\be
\label{fs}
f_s(q^2) = f_s(0) \left(1+\lambda_f q^2 \right)\,,\quad
g(q^2) = g(0) \left(1+\lambda_g q^2 \right)\,,\quad
q^2 = {(s_\pi-4 m_\pi^2)}/{(4m_\pi^2)}\,.
\ee
with $\lambda_f = \lambda_g$. Furthermore, the $f_p$ value was compatible
with zero.
The results obtained, with $\sin\theta_C=0.22$, are
\be
\label{experiment}
f_s(0) = 5.59\pm0.14\,,\quad
g(0) = 4.77\pm0.27\,,\quad
\lambda_f = 0.08\pm0.02\,.
\ee
Using isospin the result of \cite{Makoff} is $g(0) = 5.50\pm0.50$
which using PDG procedures \cite{PDG} leads to a combined
\be
\label{experiment2}
g(0) = 4.93\pm0.31\,.
\ee
Note that \cite{Makoff} neglected $f_p$ which can account for up to
5\% of the value of $g(0)$.\\

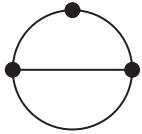
\begin{figure}
\SetScale{1.5}
\setlength{\unitlength}{1.5pt}
\begin{center}
\begin{picture}(40,40)(130,-20)
\BCirc(150,0){15}
\Line(135,0)(165,0)
\Vertex(135,0){2}
\Vertex(165,0){2}
\Vertex(150,15){2}
\end{picture}
\end{center}
\caption{\it Vertex topology. Dots refer to strong vertices or
current insertions.}
\label{figure1}
\end{figure}

{\bf 3.} As mentioned above, earlier calculations gave an indication of
potentially sizeable corrections at ${\cal O}(p^6)$.
In order to obtain the full correction we
have evaluated both form factors, $F$ and $G$, at
next-to-next-to-leading order (NNLO)
in ChPT following a diagrammatic approach. 
One faces the evaluation of
a large number of diagrams but
only one new topology, shown in Fig. \ref{figure1}, appears w.r.t. the 
vector-vector and axial-vector--axial-vector two point-functions
\cite{Amoros}.
It involves a new set
of integrals, the vertex-integrals, that can be obtained 
in terms of a two-parameter integral representation \cite{Ghinculov}.

The full evaluation leads to
rather long expressions. 
More than half of the complexity stems from the
topology of Fig. \ref{figure1} due to the large number of possible
mass and momentum combinations.
We have performed several checks on our full expressions:
$i)$ As a basic test of the algebraic programs 
we recover the one-loop expressions \cite{Kl4oneloop,BCG}.
$ii)$ All non-local divergences cancel when adding the full set 
of diagrams together with wave function renormalization. 
$iii)$ The polynomial
divergences also cancel against the counter-terms determined
in general for the even-intrinsic parity ${\cal O}(p^6)$ sector \cite{BCEinf}.
$iv)$ The double logarithms terms are in agreement with those of
\cite{BCEdble}.
$v)$ Some diagrams involving three-point one-loop integrals can
also be obtained via the renormalization of the
one-loop graphs, and hence the three-point integrals should cancel in the
final result.
$vi)$ The sunset-type integrals encountered in \cite{Ghinculov} are in 
agreement with those already calculated in \cite{Amoros} using different
methods. 
$vii)$ 
The two-particle discontinuities
of the vertex-type integrals have been checked
using the Cutkosky rules below the three-particle thresholds.
$viii)$ Our results satisfy the isospin relation 
$
M^{+-} = {M^{-0}}/{\sqrt{2}}+M^{00}.
$
In view of these checks, we trust our calculations
of the matrix-elements.

To obtain the final answer, we shift the bare quantities
 --masses and decay constants-- to the renormalized physical ones and add
the part coming from the ${\cal O}(p^6)$ Lagrangian
as determined in \cite{BCElag}.\\

{\bf 4.} One of the main problems is how to deal with the coefficients of
the ${\cal O}(p^6)$ Lagrangian. Here we follow the philosophy of most of the
other ${\cal O}(p^6)$ calculations and estimate them using resonance exchange.
This worked well at ${\cal O}(p^4)$ \cite{resonance}.
In the second paper of \cite{BTBCT} this was found to agree reasonably well
at ${\cal O}(p^6)$ for the form-factors considered there.
We use the notation as defined in \cite{Amoros} but include also terms which
did not contribute there.
Specifically we use for the vector nonet matrix $V_\mu$
\ba
{\cal L}_V &=& -\frac{1}{4}\langle V_{\mu\nu}V^{\mu\nu}\rangle
+\frac{1}{2}m_V^2\langle V_\mu V^\mu\rangle
-\frac{f_V}{2\sqrt{2}}\langle V_{\mu\nu}f_+^{\mu\nu}\rangle
-\frac{ig_V}{2\sqrt{2}}\langle V_{\mu\nu}[u^\mu,u^\nu]\rangle
+f_\chi\langle V_\mu[u^\mu,\chi_-]\rangle
\nonumber\\
&& + i \alpha_V \langle V_\mu[u_\nu,f_-^{\mu\nu}]\rangle
\ea
and for the axial-vector nonet $A_\mu$
\be
{\cal L}_A = -\frac{1}{4}\langle A_{\mu\nu}A^{\mu\nu}\rangle
+\frac{1}{2}m_A^2\langle A_\mu A^\mu\rangle
-\frac{f_A}{2\sqrt{2}}\langle A_{\mu\nu}f_-^{\mu\nu}\rangle
+\gamma_A^{(1)} \langle A_\mu u_\nu u^\mu u^\nu \rangle
+\gamma_A^{(2)} \langle A_\mu \{u^\mu, u_\nu u^\nu\}\rangle\,.
\ee
We have chosen a representation where the vectors and the axial-vectors
only start contributing to the mesonic Lagrangian at ${\cal O}(p^6)$.
For the scalar nonet we take
\begin{displaymath}
{\cal L}_S = \frac{1}{2}\langle \nabla_\mu S \nabla^\mu S \rangle
-\frac{1}{2}m_S^2\langle S^2\rangle
+ c_d \langle S u^\mu u_\mu\rangle
+ c_m \langle S \chi_+\rangle\,.
\end{displaymath}
We fixed the parameters as much as possible from experiment
or the comparison with ${\cal O}(p^4)$ \cite{pipi,resonance}.
For the remainder we use the values predicted by the
Nambu-Jona-Lasinio model \cite{Prades}.
The specific inputs used are
$m_V=$ 0.77 GeV, $m_A=$1.23 GeV, $m_S=$0.98 GeV,
$f_V=$ 0.20, $f_\chi = -$0.025, $g_V =$ 0.09, $\alpha_V=-0.014$,
$f_A =$ 0.1, $\gamma_A^{(1)}=0.006$, $\gamma_A^{(2)}=-0.01$,
$c_m=42$ MeV and $c_d=$ 32 MeV.
These are used in all the fits of Table \ref{Tableresults},
except fit 6 where we kept only
$f_V$, $f_\chi$ and $g_V$ as nonzero couplings.\\

{\bf 5.}~ChPT in the meson sector has as parameters at ${\cal O}(p^2)$
$F_0$, $B_0$ and the quark masses $\hat m = m_u=m_d$ and $m_s$.
The latter only appear multiplied with $B_0$ for a total of 3 parameters.
At ${\cal O}(p^4)$, we have $L_1^r,\ldots,L_{10}^r$ \cite{GL1} and at
${\cal O}(p^6)$ there
are 90 additional parameters \cite{BCElag}. In the two-flavour sector
the equivalent of $L_9^r$ and $L_{10}^r$ has been determined
to NNLO \cite{BTBCT}.
$L_{10}^r$ has no influence on the other quantities
considered here and the influence of $L_9^r$ is small. We will
therefore use the standard
values of \cite{GL1,BCG}.
For the ${\cal O}(p^6)$ we use
the resonance saturation approximation described above.

As input we use $F_\pi=92.4$~MeV,
$F_K/F_\pi=1.22\pm0.01$ and the physical masses of pion, kaon
and eta \cite{PDG}
in addition to $f_s(0)$, $g(0)$ and $\lambda_f$. The final input
is the ratio $m_s/\hat m$.
We further assume that the parameters $L_4^r$ and $L_6^r$ are small
because they vanish in the large $N_c$-limit.
When calculating the form-factors for $K^+\to\pi^+\pi^-e^+\nu$,
$F_\pi$ and $F_K$
we use the physical charged masses
$m_{\pi^+} = 139.56995$~MeV;
$m_{K^+} = 493.677$~MeV and $m_\eta = 547.30$~MeV.
When comparing with the quark-mass ratios we use the $\pi^0$-mass,
$m_{\pi^0} = 134.9764$~MeV, the physical $\eta$-mass
and $m_K^2 = 1/2(m_{K^+}^2+m_{K^0}^2-1.8(m_{\pi^+}^2-m_{\pi^0}^2))
= (494.53$~MeV)$^2$.
This removes the electromagnetic mass corrections including estimated
corrections to Dashen's theorem \cite{Dashen}.

In order to compare with expression (\ref{fs}) 
we have to pull out some of the kinematic behaviour encoded in the
matrix elements. We set
\ba
f_s(0) &=& F(s_\pi,s_{\ell},\cos\theta_\pi=0)\,,\quad
g(0) = G(s_\pi,s_{\ell},\cos\theta_\pi=0)\,,\nonumber\\
\lambda_f&=&\left(\frac{|F(s_\pi^\prime,s_{\ell},\cos\theta_\pi=0)|}
{|F(s_\pi,s_{\ell},\cos\theta_\pi=0)|}-1\right)
\frac{4 m_\pi^2}{s_\pi^\prime-s_\pi}\,.
\ea
We use $s_\pi= (2 m_\pi+1~\mbox{MeV})^2$ to avoid numerical
integration problems
very near threshold, $s_\pi^\prime = (336~\mbox{MeV})^2$ and $s_\ell=0$.
$f_s(0)$, $g(0)$ and $\lambda_f$ are fitted
with the experimental errors quoted in (\ref{experiment}).

$F_K/F_\pi$ is calculated with the formulas of \cite{Amoros}
and required to be $1.22\pm0.01$.
The lowest order meson masses, $m^2_{0i}$, are calculated from the
physical ones using the formulas of \cite{Amoros} and we require
$m_s/\hat m = (2m_{0K}^2-m_{0\pi}^2)/m_{0\pi}^2$
$= (3 m_{0\eta}^2-m_{0\pi}^2)/(2 m_{0\pi}^2) = 24$ with an error of 10\%.

The errors quoted above are the source of the errors quoted in our main
fit in Table \ref{Tableresults}.
They do not include an estimate of the theoretical uncertainties
due to higher orders.
The error procedure is the same as in \cite{BCG}, since we use
the same data the errors have thus not changed.

We first display our main fit and 
the canonical values. 
The ${\cal O}(p^4)$ fit illustrates the change due to
the ${\cal O}(p^6)$ corrections. It is similar
to the one-loop fits performed in \cite{BCG}, the difference
due mainly to using $L_4^r=0$ and $L_6^r=0$. 
The remainder of the table shows the variation with the choices
of input. Fit 2 is with a different choice of input quark-mass ratio,
$m_s/\hat m = 26$,
fit 3 with a different choice of $\sqrt{s_{\ell}}=0.1$~GeV,
fit 4 of $s_\pi^\prime$,
fit 5 with $L_4^r=-0.3\ 10^{-3}$ and $L_6^r=-0.2\ 10^{-3}$.
Setting $L_9^r=0$ does not change results
within the accuracy displayed. Fit 6 is  with only the
vector meson part in the resonance estimate. As can be seen 
the main contribution comes from the vector-exchange with the experimentally
determined quantities. The rest is from the scalar sector. The axial-vectors
contribute only for $s_{\ell}\ne0$. 
Fit 7 and 8 are with two different values of the renormalization scale,
 $\mu=0.5$~ GeV and $\mu=1.0$~GeV.
Our full result is $\mu$-independent
as it should be, the difference is due to the fact that we
estimated the value of the ${\cal O}(p^6)$ coefficients at the scale $\mu$.
In fit 9 we present the result with the $g(0)$ value of $\cite{Makoff}$
included.

\begin{table}[t]
\caption{\label{Tableresults}
Results for $L_i^r(\mu)$ for the various
fits described in the main text. Notice that $L_4^r, L_6^r$ and
$L_9^r$ are input, $L_4^r=0, L_6^r=0$ and  $L_9^r=6.90\  10^{-3}$.
Errors are fitting errors described in the text.
All $L_i^r(\mu)$ values quoted have been brought to the scale
$\mu=0.77$~GeV.
The standard values are $m_s/\hat m = 24$, $\sqrt{s_\pi^\prime}= 0.336$~GeV
and $s_\ell = 0$.}
\vspace{0.25cm}
\begin{small}
\begin{tabular*}{\textwidth}{*{11}{c@{\hspace{1.7mm}}}c}
\hline
               & Main Fit & \cite{GL1,BCG} &${\cal O}(p^4)$&fit 2&fit 3&
 fit 4 & fit 5 & fit 6 & fit 7 & fit 8 & fit 9\\
\hline
$10^3\,L^r_1$&   0.52$\pm$0.23&   0.37$\pm$0.23&   0.46&   0.52&   0.50&
   0.49&   0.52&   0.45&   0.42&   0.63&   0.65\\
$10^3\,L^r_2$&   0.72$\pm$0.24&   1.35$\pm$0.23&   1.49&   0.73&   0.67&
   0.74&   0.80&   0.51&   1.05&   0.73&   0.85\\
$10^3\,L^r_3$&$-$2.69$\pm$0.99&$-$3.5 $\pm$0.85&$-$3.18&$-$2.70&$-$2.57&
$-$2.73&$-$2.74&$-$2.15&$-$2.91&$-$2.68&$-$3.27\\
$10^3\,L^r_5$&   0.65$\pm$0.12&   1.4 $\pm$0.5 &   1.46&   0.63&   0.65&
   0.65&   0.82&   0.67&   0.99&   0.51&   0.60\\ 
$10^3\,L^r_7$&$-$0.31$\pm$0.15&$-$0.4 $\pm$0.2 &$-$0.49&$-$0.24&$-$0.31&
$-$0.31&$-$0.35&$-$0.30&$-$0.28&$-$0.29&$-$0.31\\
$10^3\,L^r_8$&   0.48$\pm$0.18&   0.9 $\pm$0.3 &   1.08&   0.36&   0.48&
   0.49&   0.59&   0.48&   0.52&   0.45&   0.49\\
\hline	     
changed      & & &${\cal O}(p^4)$&$m_s/\hat m$&$\sqrt{s_\ell}$&
$\sqrt{s_\pi^\prime}$&$L_4^r;L_6^r$&$V_\mu$&$\mu$&$\mu$&$g(0)$\\ 
quantity     &          &  &       & 26 & 0.1             & 0.293 
&$-0.3;-0.2$&   only&0.5&1.0&4.93\\
Unit         &          &  &       &    & GeV             & GeV   &
 $10^{-3}$&     & GeV & GeV &\\
\hline
\end{tabular*}
\end{small}
\end{table}

{\bf 6.} In conclusion, we have performed a NNLO calculation of
the $F$ and $G$ form-factors
in $K_{\ell 4}$ decays. We then used this calculation together with earlier
calculations of masses and decay constants to update
the ChPT parameters at ${\cal O}(p^4)$ with a certain range of choices
for the fitting of the available $K_{\ell 4}$ data and the estimates
of the ${\cal O}(p^6)$ parameters. None of the variations of input change the
parameters outside the experimentally determined error except
the rather extreme case of $\mu=0.5$ GeV.

\begin{figure}[htb]
\begin{center}
\epsfig{file=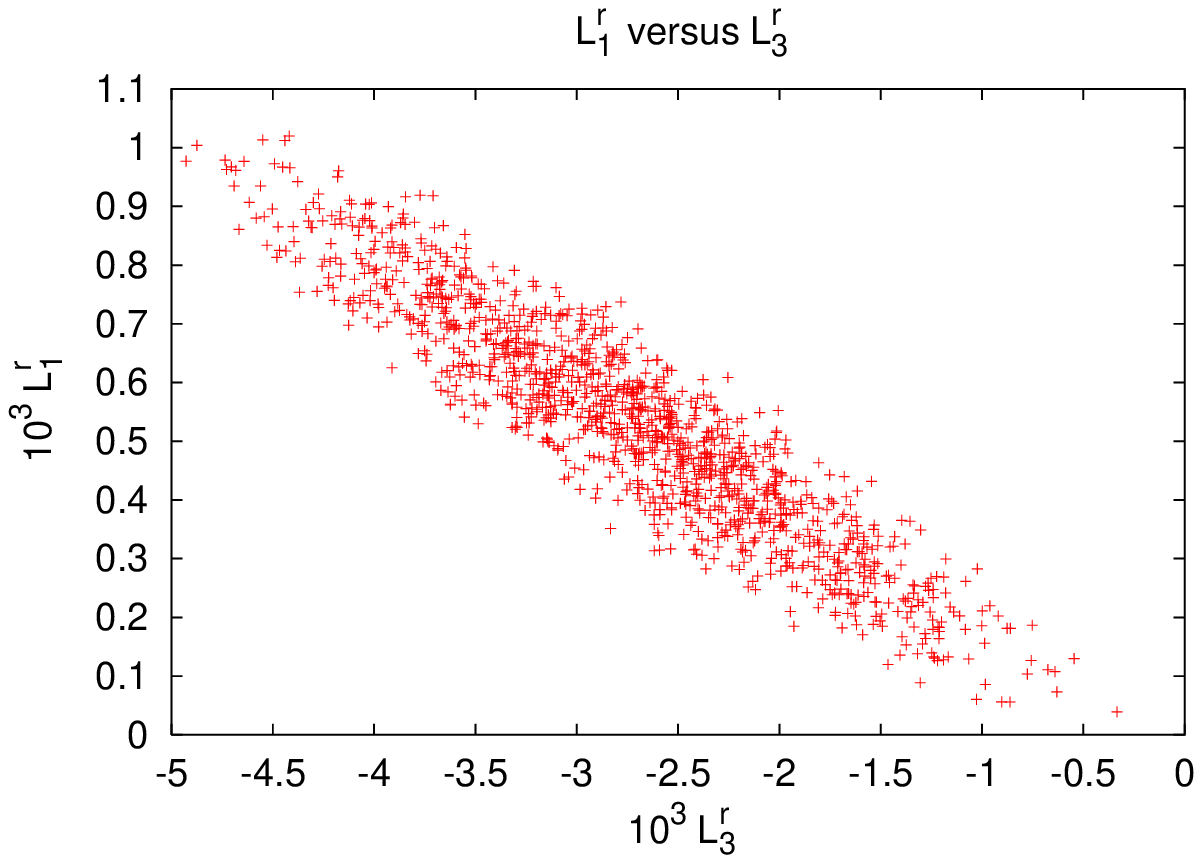,width=0.48\textwidth}
\epsfig{file=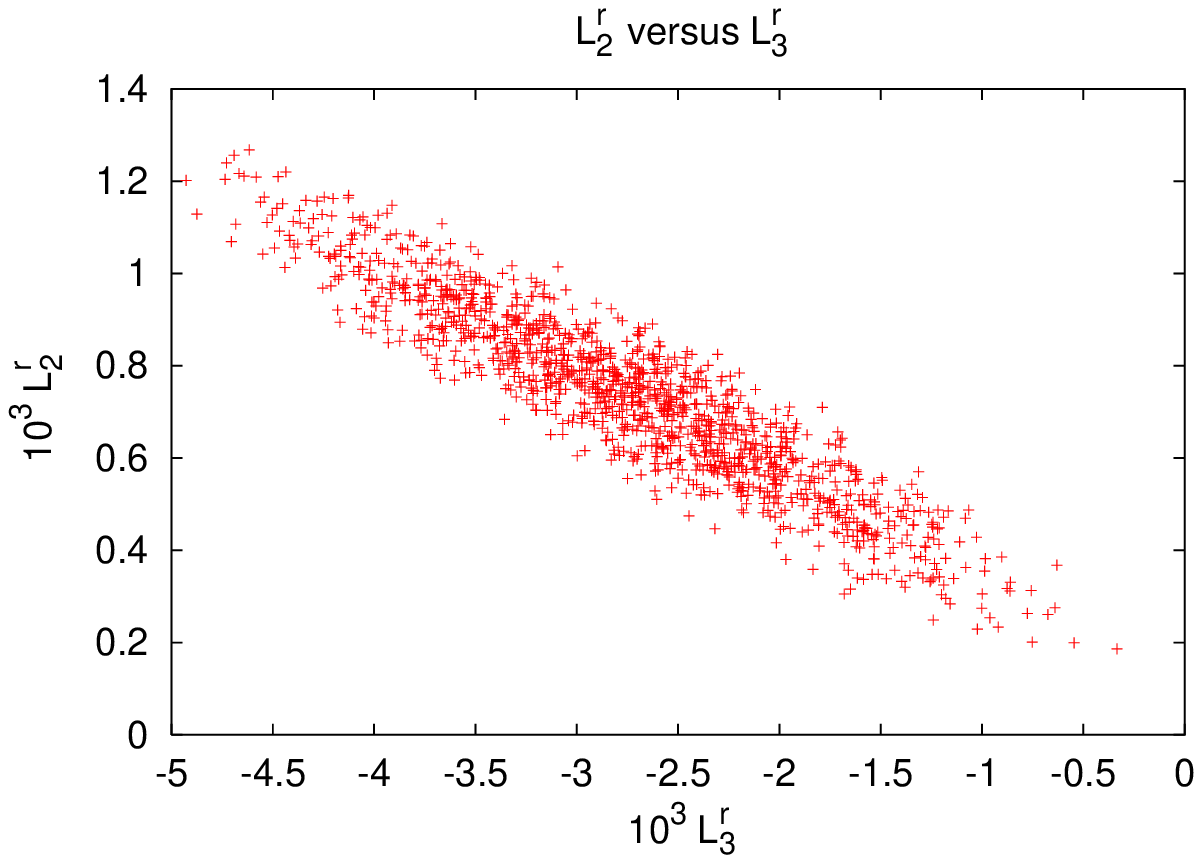,width=0.48\textwidth}
\end{center}

\caption{\label{figure2} Correlation plot of $L_1^r$ versus $L_3^r$
and $L_2^r$ versus $L_3^r$. Shown are the points from
a distribution of $L_i^r$
with the $\chi^2$ calculated from the observables described in
the text as for our main fit.
Only points within a 68\% confidence level are included.}
\end{figure}

When looking at the $K_{\ell 4}$ calculation it can be seen that
for the form-factor $F$ the estimate of higher orders of \cite{BCG}
was in the right direction. The changes are
somewhat larger than naively expected.

The errors should be taken with caution, $L_7^r$ and $L_8^r$ are strongly
anti-correlated and $L_3^r$ is strongly anti-correlated with $L_1^r$ and
$L_2^r$. As an example of this we have shown in Fig. \ref{figure2}
a distribution of sets of $L_i^r$ that fall within the
68\% confidence level limit.
If we choose as fit variables $X_1=L_2^r-2L_1^r-L_3^r$, $X_2=L_2^r$
and $X_3=(L_2^r-2L_1^r)/L_3^r$ \cite{BCG}, $X_3$ is little correlated with
the others.
We obtain
\be
X_3 = 0.12^{+0.08}_{-0.11}\,.
\ee
The large $N_c$-prediction  $|X_3|\ll 1$ is obviously well satisfied.
These questions will be discussed in more detail in a future publication.

With these values for the low-energy constants we can see how the various
quantities behave.
The numbers correspond to
 ${\cal O}(p^2)$, ${\cal O}(p^4)$ and ${\cal O}(p^6)$ contributions.
\ba
{m_{\pi^0}^2}/{m_{\pi^0}^2|_{\mbox{exp}}} &=& 0.740+0.007+0.253\,,
\nonumber\\
{m_{K}^2}/{m_{K}^2|_{\mbox{exp}}} &=& 0.689+0.024+0.287\,,
\nonumber\\
{m_{\eta}^2}/{m_{\eta}^2|_{\mbox{exp}}} &=& 0.735-0.066+0.331\,,
\nonumber\\
{F_K}/{F_\pi} &=& 1.000 + 0.135 + 0.085\,,
\nonumber\\
{F_\pi}/{F_0} &=& 1.000 +0.136-0.076\,,
\nonumber\\
f_s(0) &=&  3.78+1.15+0.66\,,
\nonumber\\
g(0) &=& 3.78+0.82+0.17\,,
\nonumber\\
\lambda_f &=& 0.000+0.127-0.047\,.
\ea
These numbers were calculated before rounding the $L_i^r$ to the number
of significant digits given in Table \ref{Tableresults}.
Contrary to what was observed in \cite{BCEdble,AB} the slope is
now mainly 
from ${\cal O}(p^4)$-effects.

Making use of the central values of the main fit in Table \ref{Tableresults} 
together with the ${\cal O}(p^4)$ relations between the two- and
three-flavour low-energy constants we estimate
$\bar l_1$ and $\bar l_2$ 
\be
\bar l_1=  0.4(-0.2)\,,\quad
\bar l_2 = 4.9(5.2)\,.
\ee
The values in brackets are those using the values of fit 9.
We remind the reader that the relation used
has possibly large ${\cal O}(p^6)$ corrections.
These should be compared respectively with $-1.7$ and $6.1$ using the $L_i^r$
from \cite{BCG} and with $\bar l_2 = 4.2$ \cite{Girlanda:1997ed}.

In conclusion, we have performed a full ${\cal O}(p^6)$ calculation
of $K_{\ell 4}$ decays. We then performed a full refit of the ${\cal O}(p^4)$
parameters of ChPT and discussed the changes and the validity of a large-$N_c$
prediction.\\

\noindent
\noindent{\bf Acknowledgements}

Work supported in part by TMR, EC--Contract No. 
ERBFMRX--CT980169 (EURODAPHNE). 
P.T. acknowledges a fellowship from the Swedish Natural Science Research 
Council.

\end{document}